\documentclass[preprint]{ptephy_v1}%%%%%% to generate preprint number

\preprintnumber{KYUSHU-HET-261, KOBE-TH-23-03} %%% %%% Insert preprint number here

\usepackage{amsmath} %for dealing with mathematics,
\usepackage{amsthm} %for dealing with theorem environments,
\usepackage{url} %It provides better support for handling and breaking URLs.

\usepackage{stmaryrd}

\DeclareMathOperator{\tr}{tr}

\newcommand{\Slash}[1]{{\ooalign{\hfil/\hfil\crcr$#1$}}}
\numberwithin{equation}{section}

\allowdisplaybreaks

\begin{document}

\title{Chiral anomaly as a composite operator in the gradient flow exact
renormalization group formalism}

%%%% To generate auto affiliation numbers please use \author{}\affil{} command

\author{Yuki Miyakawa}
\affil[1]{Department of Physics, Kyushu University, 744 Motooka, Nishi-ku,
Fukuoka 819-0395, Japan}

\author[2,3]{Hidenori Sonoda}
\affil[2]{Physics Department, Kobe University, Kobe 657-8501, Japan}
\affil[3]{Department of Physics and Astronomy, The University of Iowa,
Iowa City, Iowa 52242, USA\footnote{Visiting Scholar till 28 February 2024}}

\author[1]{Hiroshi Suzuki}

%% \author{Insert second author name here}
%% \affil{Insert second author address here}

%% \author{Insert third author name here}
%% \author[3]{Insert fourth author name here} %%% Use optional bracket [3] to change the respective address
%% \affil{Insert third author address here}

%% \author{Insert last author name here\thanks{These authors contributed equally to this work}}
%% \affil{Insert last author address here}

%%% To include the collaborator name... Please use the command "\collaborator"
%%% For example: \collaborator{ATLAS Collaboration}

\begin{abstract}%
The gradient flow exact renormalization group (GFERG) is an idea that
incorporates gauge invariant gradient flows into the formalism of the exact
renormalization group (ERG). GFERG introduces a Wilson action with a cutoff
while keeping vector gauge invariance manifestly. The details of the formalism
are still to be worked out. In this paper, we apply GFERG to construct the
Wilson action of massless Dirac fermions under the background chiral gauge
fields. By formulating the chiral anomaly as a ``composite operator,'' we make
the scale invariance of the anomaly manifest. We argue that the same
result extends to QCD.
\end{abstract}

\subjectindex{B05, B32}

\maketitle

\section{Introduction}
\label{sec:1}
The gradient flow has been introduced to lattice gauge theories as an
alternative to the renormalization group transformation~\cite{%
Narayanan:2006rf,Luscher:2009eq,Luscher:2010iy,Luscher:2011bx,Luscher:2013cpa,
Luscher:2013vga,Makino:2018rys,Carosso:2018bmz,Carosso:2019qpb}. It has already
established
itself as a practical tool to calibrate the physical scale of lattice
simulations (see Ref~\cite{dEnterria:2022hzv} for a recent review), but its
relation to the renormalization group transformation has not yet been fully
understood. What we call the gradient flow exact renormalization group
(GFERG)~\cite{%
Sonoda:2020vut,Miyakawa:2021hcx,Miyakawa:2021wus,Sonoda:2022fmk,%
Miyakawa:2022qbz} is an attempt to construct a Wilson action~\cite{%
Wilson:1973jj,
Morris:1993qb,
Becchi:1996an,
Pawlowski:2005xe,
Igarashi:2009tj,
Rosten:2010vm,
Dupuis:2020fhh} whose cutoff dependence is governed by the gradient
flow;\footnote{Articles closely related to the GFERG are~Refs.~\cite{%
Sonoda:2019ibh,Matsumoto:2020lha,Abe:2022smm}.} even the formalism is still at
an infant stage, however, and has been constructed in some details only for QED
so far.

In the GFERG formalism, we work with continuous fields in a continuous space.
The role of a physical cutoff is played by the diffusion time. As in lattice
gauge theories, the vector gauge invariance is manifestly preserved in GFERG.
With the advantage of a continuum language, we remain hopeful that the
formalism would give us new insights into the renormalization group flows
among the gauge theories.

The finite momentum cutoff of a Wilson action often leads to a misconception
that the action may miss the physics at short distances, such as the axial or
chiral anomalies. In fact the interaction vertices of a Wilson action result
from the physics at distances shorter than the cutoff distance (the inverse of
the momentum cutoff), and the Wilson action keeps the anomalies intact under
the lowering of the momentum cutoff. The axial anomaly of QED and the chiral
anomaly of the free massless fermions under the external chiral gauge fields
have been calculated in the conventional exact renormalization group (ERG)
formalism~\cite{Igarashi:2009tj,Igarashi:2011xs,Sonoda:2020gqc}.

The biggest advantage in calculating anomalies using a finite cutoff theory is
the absence of any subtlety in the calculations. There is no need to take the
momentum cutoff to infinity, and the calculations tend to be straightforward.
In this paper, we would like to consider the massless fermions under the
external chiral gauge fields, and derive the chiral anomalies using the GFERG
formalism.

Two notions play important roles: Composite operators and their scaling
dimensions. Both are standard notions in quantum field theory, but they carry
specific meanings in the ERG and GFERG formalisms. A composite
operator~\cite{Becchi:1996an} is a functional of fields that can be considered
as an infinitesimal change of the exponentiated Wilson action. If $S_\tau$ is
the Wilson action at a logarithmic distance scale~$\tau$, and
$\mathcal{O}_\tau$ a composite operator with scaling dimension~$-y$, then
$e^{y\tau}\mathcal{O}_\tau$ satisfies the same ERG or GFERG equation
as~$e^{S_\tau}$.

We will formulate the chiral anomaly as a composite operator that generates the
chiral transformation of the fermion and external gauge fields. Its
non-vanishing results in the anomalous Ward--Takahashi identities. We show that
the anomaly is a composite operator with zero scaling dimension, meaning that
the anomaly, inherited from the short distances, is independent of the scale.
This is reminiscent of Zee's proof~\cite{Zee:1972zt} of the
non-renormalization of the axial anomaly in
QED~\cite{Adler:1969er}.\footnote{The anomaly we consider is the so-called
't~Hooft anomaly~\cite{tHooft:1979rat} and its independence on the
renormalization scale has been studied
in~Refs.~\cite{Frishman:1980dq,Coleman:1982yg} in the conventional
formulation.}

In the GFERG formalism, we diffuse not only the chiral fermion fields but also
the external chiral gauge fields. Since the anomaly is scale invariant, it is
more natural to express the anomaly in terms of the undiffused gauge fields
which we call the $-1$~variables. The advantage of GFERG is the manifest
invariance under the vectorial gauge transformations. Hence, the chiral anomaly
is given most naturally in what is known as the Bardeen
form~\cite{Bardeen:1969md}.

The present paper is organized as follows. In~Sect.~\ref{sec:2}, we introduce
GFERG for chiral fermions interacting only with external chiral gauge fields.
In~Sect.~\ref{sec:3}, we introduce the chiral BRST transformation and the
chiral anomaly. In particular, we show that the anomaly is a composite operator
whose scale dimension vanishes. In~Sect.~\ref{sec:4}, we introduce the
one-particle-irreducible (1PI) formalism and use it to calculate the anomaly.
Our calculation is based upon the previous work by one of
us~\cite{Miyakawa:2022qbz}. After commenting on the chiral anomaly in QCD
in~Sect.~\ref{sec:5}, we conclude the paper in~Sect.~\ref{sec:6}. We have
prepared two appendices. In~Appendix~\ref{sec:A}, we introduce the
$-1$~variables which are the undiffused fields corresponding to the bare fields
at short distances. In~Appendix~\ref{sec:B}, we derive useful identities for
functional integrals over fermions.

Throughout the paper, we work in the $D=4$ dimensional Euclidean space, but we
keep writing $D$ to make apparent the dependence on space dimensions. We use
the following notation for momentum integrals:
\begin{equation}
   \int_p\equiv\int\frac{d^Dp}{(2\pi)^D},\qquad
   \delta(p)\equiv(2\pi)^D\delta^{(D)}(p).
\end{equation}

\section{GFERG for the free massless Dirac fermion}
\label{sec:2}
We consider $N$ free massless Dirac fermions whose classical global symmetry is
$\mathrm{U}(N)_L\times\mathrm{U}(N)_R$. To consider the anomaly associated with
the subgroup $\mathrm{SU}(N)_L\times\mathrm{SU}(N)_R$, we first introduce an
external (i.e., non-dynamical) gauge field $L_\mu^a(x)$ that couples to the
$\mathrm{SU}(N)_L$ fermion current. We postpone the introduction of the
right-handed counterpart till~Sect.~\ref{sec:4}. In the dimensionless
convention the Wilson action~$S_\tau$ is given by\footnote{This integral
representation of the Wilson action is somewhat different from that given
in~Refs.~\cite{Sonoda:2020vut,Miyakawa:2021hcx}. This representation, based on
the Gaussian integration, is more natural but equivalent.}
\begin{align}
   e^{S_\tau[\psi,\Bar{\psi},L]}
   &=\int[d\psi'd\Bar{\psi}']\,
   \exp\biggl\{
   i\int d^Dx\,
   \left[\Bar{\psi}(x)
   -e^{[(D-1)/2](\tau-\tau_0)}
   \Bar{\psi}'(t-t_0,e^{\tau-\tau_0}x)\right]
\notag\\
   &\qquad\qquad\qquad\qquad\qquad\qquad{}
   \times\left[\psi(x)
   -e^{[(D-1)/2](\tau-\tau_0)}
   \psi'(t-t_0,e^{\tau-\tau_0}x)\right]
   \biggr\}
\notag\\
   &\qquad{}
   \times
   (\Hat{s}')^{-1}e^{S_{\tau_0}[\psi',\Bar{\psi}',L']}.
\label{eq:(2.1)}
\end{align}
Here, $\tau$ is a logarithmic renormalization scale, and this
``integral representation'' relates two Wilson actions at RG scales
$\tau$ and~$\tau_0$, respectively. What we call an ``unscrambler'' is
defined by
\begin{equation}
   \Hat{s}^{-1}\equiv
   \exp\left[i\int d^Dx\,
   \frac{\overrightarrow{\delta}}{\delta\psi(x)}
   \frac{\overrightarrow{\delta}}{\delta\Bar{\psi}(x)}\right]
   =\exp\left[-i\int d^Dx\,
   \frac{\overleftarrow{\delta}}{\delta\psi(x)}
   \frac{\overrightarrow{\delta}}{\delta\Bar{\psi}(x)}\right].
\label{eq:(2.2)}
\end{equation}
$(\Hat{s}')^{-1}$ is obtained by replacing $\psi, \Bar{\psi}$ by
$\psi', \Bar{\psi}'$. In~Eq.~\eqref{eq:(2.1)}, the Wilson action is
obtained by the ``blocking procedure'' implemented by the Gaussian
integration; it is thus natural not to include the Gaussian integration for
the external gauge field. The primed field variables in the exponent
are given by the solutions to the diffusion equations,
\begin{subequations}
\label{eq:(2.3)}
\begin{align}
   \partial_{t'}\psi'(t',x)
   &=\left\{
   \left[\partial_\mu+L_\mu^{\prime a}(t',x)T^aP_L
   \right]^2
   -\alpha_0\partial_\mu L_\mu^{\prime a}(t',x)T^aP_L
   \right\}
   \psi'(t',x)
   \equiv\Delta'\psi'(t',x),
\\
   \partial_{t'}\Bar{\psi}'(t',x)
   &=\Bar{\psi}'(t',x)
   \left\{
   \left[\overleftarrow{\partial}_\mu
   -P_RL_\mu^{\prime a}(t',x)T^a
   \right]^2
   +\alpha_0P_R\partial_\mu L_\mu^{\prime a}(t',x)T^a
   \right\}
   \equiv\Bar{\psi}'(t',x)\overleftarrow{\Delta}',
\end{align}
\end{subequations}
where the diffusion time~$t'$ ranges from~$0$ to~$t-t_0$, given by the
difference of the logarithmic RG scales~$\tau-\tau_0$ as
\begin{equation}
   t-t_0=e^{2(\tau-\tau_0)}-1.
\label{eq:(2.4)}
\end{equation}
$\alpha_0$ is an arbitrary positive parameter that controls the diffusion of
the unphysical longitudinal part, $T^a$ denote anti-Hermitian generators
of~$\mathrm{SU}(N)$ (satisfying $[T^a,T^b]=f^{abc}T^c$), and
\begin{equation}
   P_L\equiv\frac{1-\gamma_5}{2},\qquad
   P_R\equiv\frac{1+\gamma_5}{2},
\label{eq:(2.5)}
\end{equation}
where $\gamma_5\equiv\gamma_1\gamma_2\gamma_3\gamma_4$. The initial conditions
for the above diffusion equations are given by the integration variables
in~Eq.~\eqref{eq:(2.1)}, i.e.,
\begin{equation}
   \psi'(0,x)=\psi'(x),\qquad
   \Bar{\psi}'(0,x)=\Bar{\psi}'(x).
\label{eq:(2.6)}
\end{equation}

In Eqs.~\eqref{eq:(2.3)}, the external gauge field also satisfies the diffusion
equation
\begin{equation}
   \partial_{t'} L_\mu^{\prime a}(t',x)
   =D_{L\nu}'F_{L\nu\mu}^{\prime a}(t',x)
   +\alpha_0D_{L\mu}'\partial_\nu L_\nu^{\prime a}(t',x)
   \equiv\Delta'L_\mu^{\prime a}(t',x),
\label{eq:(2.7)}
\end{equation}
where
\begin{equation}
   F_{L\mu\nu}^{\prime a}(t',x)
   \equiv
   \partial_\mu L_\nu^{\prime a}(t',x)-\partial_\nu L_\mu^{\prime a}(t',x)
   +f^{abc}L_\mu^{\prime b}(t',x)L_\nu^{\prime c}(t',x)
\label{eq:(2.8)}
\end{equation}
is the field strength, and the covariant derivative is defined by
\begin{equation}
   D_{L\mu}'X^{\prime a}(t',x)
   \equiv
   \partial_\mu X^{\prime a}(t',x)
   +f^{abc}L_\mu^{\prime b}(t',x)X^{\prime c}(t',x).
\label{eq:(2.9)}
\end{equation}
We select a particular solution satisfying the boundary condition
\begin{equation}
   L_\mu^a(x)=e^{\tau-\tau_0}L_\mu^{\prime a}(t-t_0,e^{\tau-\tau_0}x).
\label{eq:(2.10)}
\end{equation}
This determines
\begin{equation}
   L_\mu^{\prime a}(x)=L_\mu^{\prime a}(0,x)
\label{eq:(2.11)}
\end{equation}
as a functional of~$L_\mu^a(x)$.

The Wilson action $S_\tau$ is constructed so that the correlation functions,
modified by the insertion of~$\hat{s}^{-1}$, are given by~\cite{Sonoda:2015bla}
\begin{align}
   &\left\langle
   \hat{s}^{-1}\psi(x_1)\dotsb\psi(x_n)
   \Bar{\psi}(y_1)\dotsb\Bar{\psi}(y_n)\right\rangle_{S_\tau}
\notag\\
   &=e^{2n[(D-1)/2](\tau-\tau_0)}
   \left\langle
   \hat{s}^{-1}\psi(t-t_0,e^{\tau-\tau_0}x_1)
   \dotsb\Bar{\psi}(t-t_0,e^{\tau-\tau_0}y_1)\dotsb
   \right\rangle_{S_{\tau_0}}.
\label{eq:(2.12)}
\end{align}
(We derive this in~Appendix~\ref{sec:B}.) Hence, differentiating this with
respect to~$\tau$, we obtain
\begin{align}
   &\partial_\tau
   \left\langle
   \hat{s}^{-1}\psi(x_1)\dotsb\psi(x_n)
   \Bar{\psi}(y_1)\dotsb\Bar{\psi}(y_n)
   \right\rangle_{S_\tau}
\notag\\
   &\qquad{}
   +\int d^Dx\,
   \left(1+x\cdot\partial+2\Delta\right)
   L_\mu^a(x)\cdot\frac{\delta}{\delta L_\mu^a(x)}
   \left\langle\hat{s}^{-1}\psi(x_1)\dotsb\Bar{\psi}(y_n)\right\rangle_{S_\tau}
\notag\\
   &=\sum_{i=1}^n\left(\frac{D-1}{2}+x_i\cdot\partial_{x_i}+2\Delta_{x_i}\right)
   \left\langle
   \hat{s}^{-1}\psi(x_1)\dotsb\Bar{\psi}(y_n)\right\rangle_{S_\tau}
\notag\\
   &\qquad{}
   +\sum_{i=1}^n
   \left\langle\hat{s}^{-1}\psi(x_1)\dotsb\Bar{\psi}(y_n)\right\rangle_{S_\tau}
   \left(\frac{D-1}{2}+\overleftarrow{\partial}_{y_i}\cdot y_i
   +2\overleftarrow{\Delta}_{y_i}\right).
\label{eq:(2.13)}
\end{align}
This amounts to the GFERG equation given by
\begin{align}
   &\left[
   \frac{\partial}{\partial\tau}
   +\int d^Dx\,
   \left(1+x\cdot\partial+2\Delta\right)L_\mu^a(x)\cdot
   \frac{\delta}{\delta L_\mu^a(x)}
   \right]
   e^{S_\tau}
\notag\\
   &=\int d^Dx\,
   \tr\left\{\left(\frac{D-1}{2}+x\cdot\partial+2\Delta\right)
   \left[
   \psi(x)+i\frac{\overrightarrow{\delta}}{\delta\Bar{\psi}(x)}
   \right]
   e^{S_\tau}\right\}
   \frac{\overleftarrow{\delta}}{\delta\psi(x)}
\notag\\
   &\qquad{}
   +\int d^Dx\,
   \tr
   \frac{\overrightarrow{\delta}}{\delta\Bar{\psi}(x)}
   \left\{e^{S_\tau}
   \left[\Bar{\psi}(x)+i\frac{\overleftarrow{\delta}}{\delta\psi(x)}\right]
   \left(
   \frac{D-1}{2}+\overleftarrow{\partial}\cdot x+2\overleftarrow{\Delta}
   \right)\right\},
\label{eq:(2.14)}
\end{align}
where the ``scrambler'' $\Hat{s}$, the inverse of $\hat{s}^{-1}$, is given by
reversing the sign in the exponent in~Eq.~\eqref{eq:(2.2)} as
\begin{equation}
   \hat{s}\equiv
   \exp\left[
   -i\int d^Dx\,\frac{\overrightarrow{\delta}}{\delta\psi(x)}
   \frac{\overrightarrow{\delta}}{\delta\Bar{\psi}(x)}
   \right]
   =\exp\left[i\int d^Dx\,
   \frac{\overleftarrow{\delta}}{\delta\psi(x)}
   \frac{\overrightarrow{\delta}}{\delta\Bar{\psi}(x)}
   \right],
\label{eq:(2.15)}
\end{equation}
and $\Delta$'s are defined by the expressions in~Eqs.~\eqref{eq:(2.3)}
and~\eqref{eq:(2.7)} by removing all the primes.

In this paper, we assume that the Wilson action be at most bi-linear in the
fermion fields. Under~Eq.~\eqref{eq:(2.14)}, this property is preserved. As
discussed in~Refs.~\cite{Sonoda:2020vut,Miyakawa:2021hcx} and as we explain in
what follows, the above diffusion equations (referred to as gradient flow
equations) commute with the external gauge transformations. On the other hand,
the Gaussian integration and the unscrambler preserve only the \emph{vectorial
part\/} $\mathrm{SU}(N)_V$ of the external gauge invariance. We can thus take
it for granted that the Wilson action is invariant under~$\mathrm{SU}(N)_V$.

\section{Chiral anomaly}
\label{sec:3}
\subsection{BRST transformation and the chiral anomaly}
To analyze possible non-invariance of the Wilson action under the gauge
transformations of~$L_\mu^a(x)$, we introduce a Grassmann-odd ghost
field~$\chi_L^a(x)$, which simply plays the role of a transformation function
of the gauge transformations. The chiral BRST transformation, corresponding to
an arbitrary $\mathrm{SU}(N)_L$ gauge transformation, is generated by the
differential operator,
\begin{align}
   \Hat{\delta}_L
   &\equiv
   \int d^Dx\,\biggl\{
   -\chi_L^a(x)T^aP_L\psi(x)\frac{\delta}{\delta\psi(x)}
   +\chi_L^a(x)\Bar{\psi}(x)P_RT^a\frac{\delta}{\delta\Bar{\psi}(x)}
\notag\\
   &\qquad\qquad\qquad{}
   +\left[\partial_\mu\chi_L^a(x)+f^{abc}L_\mu^b(x)\chi_L^c(x)\right]
   \frac{\delta}{\delta L_\mu^a(x)}
   -\frac{1}{2}f^{abc}\chi_L^b(x)\chi_L^c(x)
   \frac{\delta}{\delta\chi_L^a(x)}
   \biggr\},
\label{eq:(3.1)}
\end{align}
which is nilpotent~$\Hat{\delta}_L^2=0$ by construction. Using the
unscrambler~\eqref{eq:(2.2)}, we further define
\begin{equation}
   \Tilde{\Hat{\delta}}_L
   \equiv
   \Hat{s}\Hat{\delta}_L\Hat{s}^{-1},
\label{eq:(3.2)}
\end{equation}
which is still nilpotent:
\begin{equation}
   \Tilde{\Hat{\delta}}_L^2=0.
\label{eq:(3.3)}
\end{equation}
We define the chiral anomaly---possible non-invariance of the Wilson
action~$S_\tau$ under the BRST transformation---by
\begin{align}
   \mathcal{Q}_{L\tau}
   &\equiv e^{-S_\tau}\Tilde{\Hat{\delta}}_Le^{S_\tau}
   =e^{-S_\tau}\Hat{s}\Hat{\delta}_L\left(\Hat{s}^{-1}e^{S_\tau}\right)
\notag\\
   &=\int d^Dx\,
   \Biggl(
   \left[\partial_\mu\chi_L^a(x)+f^{abc}L_\mu^b(x)\chi_L^c(x)\right]
   \frac{\delta S_\tau}{\delta L_\mu^a(x)}
\notag\\
   &\qquad\qquad\qquad{}
   +\tr\left\{\chi_L^aT^aP_L
   \left[\psi(x)+i\frac{\overrightarrow{\delta}}{\delta\Bar{\psi}(x)}\right]
   e^{S_\tau}\right\}\frac{\overleftarrow{\delta}}{\delta\psi(x)}e^{-S_\tau}
\notag\\
   &\qquad\qquad\qquad{}
   -e^{-S_\tau}\tr\frac{\overrightarrow{\delta}}{\delta\Bar{\psi}(x)}
   \left\{e^{S_\tau}
   \left[\Bar{\psi}(x)+i\frac{\overleftarrow{\delta}}{\delta\psi(x)}\right]
   P_R\chi_L^a(x)T^a\right\}
   \Biggr).
\label{eq:(3.4)}
\end{align}
Its correlation functions satisfy
\begin{align}
   &\left\langle
   \mathcal{Q}_{L\tau}\Hat{s}^{-1}
   \psi(x_1)\dotsb\Bar{\psi}(y_1)\dotsb\right\rangle_{S_\tau}
\notag\\
   &=\int[d\psi d\Bar{\psi}]\,e^{S_\tau}\mathcal{Q}_{L\tau}\cdot
   \hat{s}^{-1}\left[\psi(x_1)\dotsb\Bar{\psi}(y_1)\dotsb\right]
\notag\\
   &=\int[d\psi d\Bar{\psi}]\,
   \hat{\delta}_L
   \left(\hat{s}^{-1}e^{S_\tau}\right)
   \left[\psi(x_1)\dotsb\Bar{\psi}(y_1)\dotsb\right]
\notag\\
   &=\int d^Dx\,
   \Biggl\{
   \left[\partial_\mu\chi_L^a(x)+f^{abc}L_\mu^b(x)\chi_L^c(x)\right]
   \frac{\delta}{\delta L_\mu^a(x)}
   \left\langle
   \Hat{s}^{-1}\psi(x_1)\dotsb\Bar{\psi}(y_1)\dotsb
   \right\rangle_{S_\tau}
\notag\\
   &\qquad\qquad\qquad{}
   +\sum_i
   \chi_L^a(x_i)
   \left\langle
   \Hat{s}^{-1}\psi(x_1)\dotsb
   T^aP_L\psi(x_i)\dotsb
   \Bar{\psi}(y_1)\dotsb
   \right\rangle_{S_\tau}
\notag\\
   &\qquad\qquad\qquad{}
   -\sum_i
   \chi_L^a(y_i)
   \left\langle
   \Hat{s}^{-1}
   \psi(x_1)\dotsb\Bar{\psi}(y_1)\dotsb
   \Bar{\psi}(y_i)P_RT^a \dotsb
   \right\rangle_{S_\tau}
   \Biggr\},
\label{eq:(3.5)}
\end{align}
where $\left\langle\Hat{s}^{-1}\psi(x_1)\dotsb\Bar{\psi}(y_1)\dotsb%
\right\rangle_{S_\tau}$ is the modified correlation function satisfying
Eq.~\eqref{eq:(2.13)}. If the theory is gauge invariant, the right-hand side
should vanish, implying the absence of anomaly~$\mathcal{Q}_{L\tau}=0$. If
$\mathcal{Q}_{L\tau}\neq0$, Eq.~\eqref{eq:(3.5)} amounts to an anomalous
Ward--Takahashi identity. Assuming that the Wilson action~$S_\tau$ is local,
$\mathcal{Q}_{L\tau}$ is also a local operator. Furthermore, from the nilpotency
of~$\tilde{\hat{\delta}}_L$, the chiral anomaly satisfies the Wess--Zumino (WZ)
consistency condition,
\begin{equation}
   e^{-S_\tau}\Tilde{\Hat{\delta}}_Le^{S_\tau}\mathcal{Q}_{L\tau}
   =e^{-S_\tau}\tilde{\hat{\delta}}_L^2e^{S_\tau}
   =0.
\label{eq:(3.6)}
\end{equation}

\subsection{Chiral anomaly as a scaling composite operator}
The chiral anomaly $\mathcal{Q}_{L\tau}$ is linear in~$\chi_L^a$. In order to
promote the anomaly to a scaling composite operator, we need to diffuse the
ghost field as
\begin{equation}
   \partial_{t'}\chi_L^{\prime a}(t',x)
   =\alpha_0D_{L\mu}'\partial_\mu\chi_L^{\prime a}(t',x)
   \equiv\Delta'\chi_L^{\prime a}(t',x).
\label{eq:(3.7)}
\end{equation}
We identify the original ghost field $\chi_L^a(x)$ with the boundary condition
for the above diffusion equation
\begin{align}
   \chi_L^a(x)=\chi_L^{\prime a}(t-t_0,e^{\tau-\tau_0}x).
\label{eq:(3.8)}
\end{align}
This is analogous to~Eq.~\eqref{eq:(2.7)}. Under this setup, the
anomaly~\eqref{eq:(3.4)} behaves in a very simple way under the GFERG
transformation.

To see this, let us apply the BRST transformation,
\begin{subequations}
\label{eq:(3.9)}
\begin{align}
   \delta_L\psi'(t',x)
   &=-\chi_L^{\prime a} (t', x)T^aP_L\psi(t',x),\qquad
   \delta_L\Bar{\psi}(t', x)
   =\chi_L^{\prime a} (t', x)\Bar{\psi}(t', x)P_RT^a,
\\
   \delta_LL_\mu^{\prime a}(t',x)
   &=\partial_\mu\chi_L^{\prime a}(t',x)
   +f^{abc}L_\mu^{\prime b}(t',x)\chi_L^{\prime c}(t',x),
\\
   \delta_L\chi_L^{\prime a}(t',x)
   &=-\frac{1}{2}f^{abc}\chi_L^{\prime b}(t',x)\chi_L^{\prime c}(t',x),
\end{align}
\end{subequations}
on the diffusion equations given by~Eqs.~\eqref{eq:(2.3)}, \eqref{eq:(2.7)},
and~\eqref{eq:(3.7)}. After some calculations, we find that the changes in the
diffusion equations are proportional
to~$(\partial_{t'}-\alpha_0D_{L\mu}'\partial_\mu)\chi_L^{\prime a}(t',x)$ which
vanishes thanks to~Eq.~\eqref{eq:(3.7)}. Therefore, with the adoption
of~Eq.~\eqref{eq:(3.7)} for~$\chi_L^{\prime a}$, all the diffusion equations are
invariant under the BRST transformation~\eqref{eq:(3.9)}. In other words, the
diffusion and the BRST transformation commute.

The BRST transformation~\eqref{eq:(3.9)} acts not only at the intermediate~$t'$
but also at the boundaries $t'=0$ and~$t'=t-t_0$. We wish to show that the
consistency of the diffusion with~Eq.~\eqref{eq:(3.9)} implies
\begin{align}
   &\mathcal{Q}_{L \tau}e^{S_\tau[\psi,\Bar{\psi},L]}
\notag\\
   &=\int[d\psi'd\Bar{\psi}']\,
   \exp\biggl\{
   i\int d^Dx\,
   \left[
   \Bar{\psi}(x)-e^{[(D-1)/2](\tau-\tau_0)}\Bar{\psi}'(t-t_0,e^{\tau-\tau_0}x)
   \right]
\notag\\
   &\qquad\qquad\qquad\qquad\qquad\qquad{}
   \times
   \left[
   \psi(x)-e^{[(D-1)/2](\tau-\tau_0)}\psi'(t-t_0,e^{\tau-\tau_0}x)\right]
   \biggr\}
\notag\\
   &\qquad{}
   \times
   (\Hat{s}')^{-1}\mathcal{Q}_{L\tau_0}'
   e^{S_{\tau_0}[\psi',\Bar{\psi}',L']},
\label{eq:(3.10)}
\end{align}
where $L_\mu^a$ on the left-hand side is related to~$L_\mu^{\prime a}$ on the
right-hand side by the diffusion equation as Eqs.~\eqref{eq:(2.10)}
and~\eqref{eq:(2.11)}. For brevity we merely sketch a (rather formal)
derivation.

Following the definition \eqref{eq:(3.4)} we obtain
\begin{equation}
   \mathcal{Q}_{L\tau}e^{S_\tau[\psi,\Bar{\psi},L]}
   =\Tilde{\Hat{\delta}}_Le^{S_\tau[\psi,\Bar{\psi},L]} 
   =\Hat{s}\Hat{\delta}_L\Hat{s}^{-1}e^{S_\tau[\psi,\Bar{\psi},L]}.
\label{eq:(3.11)}
\end{equation}
Applying $\hat{s}^{-1}$ on~Eq.~\eqref{eq:(2.1)} and using the
identity~\eqref{eq:(B6)} obtained in~Appendix~\ref{sec:B}, we obtain
\begin{align}
   &\hat{s}^{-1}e^{S_\tau[\psi,\Bar{\psi},L]}
\notag\\
   &=\hat{s}^{-1}
   \int[d\psi'd\Bar{\psi}']\,
   \exp\biggl\{
   i\int d^Dx\,
   \left[
   \Bar{\psi}(x)
   -e^{[(D-1)/2](\tau-\tau_0)}\Bar{\psi}'(t-t_0,e^{\tau-\tau_0}x)
   \right]
\notag\\
   &\qquad\qquad\qquad\qquad\qquad\qquad\qquad{}
   \times
   \left[
   \psi(x)-e^{[(D-1)/2](\tau-\tau_0)}\psi'(t-t_0,e^{\tau-\tau_0}x)
   \right]\biggr\}
\notag\\
   &\qquad{}
   \times
   \left(\hat{s}'\right)^{-1}e^{S_{\tau_0}[\psi',\Bar{\psi}',L']}
\notag\\
   &=\int[d\psi'd\Bar{\psi}']\,
   \prod_x\Bigl[
   \delta\left(
   \Bar{\psi}(x)-e^{[(D-1)/2](\tau-\tau_0)}\Bar{\psi}'(t-t_0,e^{\tau-\tau_0}x)
   \right)
\notag\\
   &\qquad\qquad\qquad\qquad{}
   \times\delta\left(
   \psi(x)-e^{[(D-1)/2](\tau-\tau_0)}\psi'(t-t_0,e^{\tau-\tau_0}x)
   \right)
   \Bigr]
\notag\\
   &\qquad{}
   \times
   \left(\hat{s}'\right)^{-1}e^{S_{\tau_0}[\psi',\Bar{\psi}',L']}.
\label{eq:(3.12)}
\end{align}
Since the BRST transformation commutes with diffusion, we obtain
\begin{equation}
   \left(\hat{\delta}_L+\hat{\delta}'_L\right)
   \prod_x\left[
   \delta\left(\psi(x)-\dotsb\right)\cdot
   \delta\left(\Bar{\psi}(x)-\dotsb\right)
   \right]=0.
\label{eq:(3.13)}
\end{equation}
We then obtain
\begin{align}
   &\hat{\delta}_L\hat{s}^{-1}e^{S_\tau[\psi,\Bar{\psi},L]}
\notag\\
   &=\int[d\psi'd\Bar{\psi}']\,
   \left(-\hat{\delta}_L'\right)\prod_x\Bigl[
   \delta\left(
   \Bar{\psi}(x)-e^{[(D-1)/2](\tau-\tau_0)}\Bar{\psi}'(t-t_0,e^{\tau-\tau_0}x)
   \right)
\notag\\
   &\qquad\qquad\qquad\qquad{}
   \times\delta\left(
   \psi(x)-e^{[(D-1)/2](\tau-\tau_0)}\psi'(t-t_0,e^{\tau-\tau_0}x)
   \right)
   \Bigr]
\notag\\
   &\qquad{}
   \times
   \left(\hat{s}'\right)^{-1}e^{S_{\tau_0}[\psi',\Bar{\psi}',L']}
\notag\\
   &=\int[d\psi'd\Bar{\psi}']\,
   \prod_x\Bigl[
   \delta\left(
   \Bar{\psi}(x)-e^{[(D-1)/2](\tau-\tau_0)}\Bar{\psi}'(t-t_0,e^{\tau-\tau_0}x)
   \right)
\notag\\
   &\qquad\qquad\qquad\qquad{}
   \times\delta\left(
   \psi(x)-e^{[(D-1)/2](\tau-\tau_0)}\psi'(t-t_0,e^{\tau-\tau_0}x)
   \right)
   \Bigr]
\notag\\
   &\qquad{}
   \times
   \hat{\delta}_L'
   \left(\hat{s}'\right)^{-1}e^{S_{\tau_0}[\psi',\Bar{\psi}',L']}
\notag\\
   &=\hat{s}^{-1}
   \int[d\psi'd\Bar{\psi}']\,
   \exp\biggl\{
   i\int d^Dx\,
   \left[
   \Bar{\psi}(x)
   -e^{[(D-1)/2](\tau-\tau_0)}\Bar{\psi}'(t-t_0,e^{\tau-\tau_0}x)
   \right]
\notag\\
   &\qquad\qquad\qquad\qquad\qquad\qquad\qquad{}
   \times
   \left[
   \psi(x)-e^{[(D-1)/2](\tau-\tau_0)}\psi'(t-t_0,e^{\tau-\tau_0}x)
   \right]\biggr\}
\notag\\
   &\qquad{}
   \times
   \hat{\delta}_L'
   \left(\hat{s}'\right)^{-1}e^{S_{\tau_0}[\psi',\Bar{\psi}',L']},
\label{eq:(3.14)}
\end{align}
where $\hat{\delta}'_L$ was moved from left to right by the integration by
parts. Finally, applying~$\hat{s}$ from the left, and using
Eq.~\eqref{eq:(3.11)} and
\begin{equation}
   \hat{\delta}'_L\left(\hat{s}'\right)^{-1}
   =\left(\hat{s}'\right)^{-1}\left(\hat{s}'\right)\hat{\delta}'_L
   \left(\hat{s}'\right)^{-1}
   =\left(\hat{s}'\right)^{-1}\mathcal{Q}'_{L\tau_0},
\label{eq:(3.15)}
\end{equation}
we obtain the desired result, Eq.~\eqref{eq:(3.10)}.

Comparing Eq.~\eqref{eq:(3.10)} with Eq.~\eqref{eq:(2.1)}, we see that
$\mathcal{Q}_{L\tau}e^{S_\tau}=\tilde{\hat{\delta}}_Le^{S_\tau}$ satisfies the same
GFERG equation as~$e^{S_\tau}$. To be more precise, using an infinitesimal
Grassmann-odd $\eta$, we construct
\begin{equation}
   e^{S_\tau}+\eta\Tilde{\Hat{\delta}}_Le^{S_\tau}
   =e^{S_\tau}(1+\eta\mathcal{Q}_{L\tau})
   =e^{S_\tau+\eta\mathcal{Q}_{L\tau}}.
\label{eq:(3.16)}
\end{equation}
This satisfies the same GFERG equation as~$e^{S_\tau}$ where the
$\tau$~evolution of the ghost field, given by Eqs.~\eqref{eq:(3.7)}
and~\eqref{eq:(3.8)}, is supplemented.

In the general ERG framework, we call a functional $\mathcal{O}_\tau(x)$ a
composite operator with scaling dimension~$d_{\mathcal{O}\tau}$, if the Wilson
action modified infinitesimally as
\begin{equation}
   S_\tau
   +\epsilon e^{-\int^\tau d\tau'\,d_{\mathcal{O}\tau'}}\mathcal{O}_\tau(e^{-\tau}x)
\label{eq:(3.17)}
\end{equation}
still satisfies the ERG equation. Using this terminology, we conclude that the
chiral anomaly~$\mathcal{Q}_{L\tau}$~\eqref{eq:(3.4)}, with no coordinate
dependence, is a composite operator with scaling dimension~$0$.\footnote{A
construction of the anomaly analogous to the present one has been studied
in~Ref.~\cite{Sonoda:2020gqc} in the context of the conventional ERG, where the
vectorial gauge symmetry is not manifestly preserved.}

Since $e^{S_\tau}+e^{S_\tau}\eta\mathcal{Q}_{L\tau}$ and~$e^{S_\tau}$ satisfy the
same GFERG equation~\eqref{eq:(2.14)}, we obtain the GFERG equation
for~$\mathcal{Q}_{L\tau}$ as
\begin{align}
   &\frac{\partial}{\partial\tau}\mathcal{Q}_{L\tau}
\notag\\
   &=\int d^Dx\,
   \tr\left\{\left(\frac{D-1}{2}+x\cdot\partial+2\Delta\right)
   \left[
   \psi(x)+i\frac{\overrightarrow{\delta}}{\delta\Bar{\psi}(x)}
   \right]
   \left(\mathcal{Q}_{L\tau}e^{S_\tau}\right)\right\}
   \frac{\overleftarrow{\delta}}{\delta\psi(x)}e^{-S_\tau}
\notag\\
   &\qquad\qquad\qquad{}
   +e^{-S_\tau}
   \frac{\overrightarrow{\delta}}{\delta\Bar{\psi}(x)}
   \left\{\left(\mathcal{Q}_{L\tau}e^{S_\tau}\right)
   \left[\Bar{\psi}(x)+i\frac{\overleftarrow{\delta}}{\delta\psi(x)}\right]
   \left(
   \frac{D-1}{2}+\overleftarrow{\partial}\cdot x+2\overleftarrow{\Delta}
   \right)\right\}
\notag\\
   &\qquad{}
   -\int d^Dx\,
   \left(1+x\cdot\partial+2\Delta\right)L_\mu^a(x)\cdot
   \frac{\delta}{\delta L_\mu^a(x)}
   \mathcal{Q}_{L\tau}
\notag\\
   &\qquad{}
   -\int d^Dx\,
   \left(x\cdot\partial+2\Delta\right)\chi_L^a(x)\cdot
   \frac{\delta}{\delta\chi_L^a(x)}
   \mathcal{Q}_{L\tau}
\label{eq:(3.18)}
\end{align}
where the last term arises from the the dependence of~$\mathcal{Q}_{L\tau}$ on
the ghost field. It is important that Eq.~\eqref{eq:(3.18)} is linear
in~$\mathcal{Q}_{L\tau}$. Hence, if $\mathcal{Q}_{L\tau}=0$ for a certain~$\tau$,
then $\mathcal{Q}_{L\tau}=0$ for any~$\tau$.

\section{Chiral anomaly in the 1PI formalism}
\label{sec:4}
\subsection{1PI formalism and the $-1$ variables}
\label{sec:4.1}
To study the structure of the anomaly systematically, it is convenient to
employ the 1PI formalism of GFERG~\cite{Sonoda:2022fmk}. The Legendre
transformation from the Wilson action~$S_\tau$ to the 1PI
action~${\mit\Gamma}_\tau$ is given by
\begin{align}
   &{\mit\Gamma}_\tau[\Psi,\Bar{\Psi},L]
   +i\int d^Dx\,\Bar{\Psi}(x)\Psi(x)
\notag\\
   &\equiv
   S_\tau[\psi,\Bar{\psi},L]
   -i\int d^Dx\,\Bar{\psi}(x)\psi(x)
   +i\int d^Dx\,\left[
   \Bar{\Psi}(x)\psi(x)+\Bar{\psi}(x)\Psi(x)
   \right].
\label{eq:(4.1)}
\end{align}
Extremizing the right-hand side with respect to~$\psi$ and~$\Bar{\psi}$, we
obtain
\begin{equation}
   \Psi(x)=\psi(x)
   +i\frac{\overrightarrow{\delta}}{\delta\Bar{\psi}(x)}S_\tau,\qquad
   \Bar{\Psi}(x)=\Bar{\psi}(x)
   +iS_\tau\frac{\overleftarrow{\delta}}{\delta\psi(x)}.
\label{eq:(4.2)}
\end{equation}
It follows
\begin{equation}
   \Psi(x)\frac{\overleftarrow{\delta}}{\delta\psi(y)}
   =\delta(x-y)
   +i\frac{\overrightarrow{\delta}}{\delta\Bar{\psi}(x)}S_\tau
   \frac{\overleftarrow{\delta}}{\delta\psi(y)}
   =\frac{\overrightarrow{\delta}}{\delta\Bar{\psi}(x)}\Bar{\Psi}(y).
\label{eq:(4.3)}
\end{equation}
Note that we do not Legendre transform with respect to the external gauge
field. The inverse Legendre transform is given by
\begin{equation}
   \frac{1}{i}
   \frac{\overrightarrow{\delta}}{\delta\Bar{\Psi}(x)}{\mit\Gamma}_\tau
   +\Psi(x)=\psi(x),\qquad
   \frac{1}{i}
   {\mit\Gamma}_\tau\frac{\overleftarrow{\delta}}{\delta\Psi(x)}
   +\Bar{\Psi}(x)=\Bar{\psi}(x).
\label{eq:(4.4)}
\end{equation}
Hence, we obtain
\begin{equation}
   \frac{\overrightarrow{\delta}}{\delta\Bar{\Psi}(x)}{\mit\Gamma}_\tau
   =\frac{\overrightarrow{\delta}}{\delta\Bar{\psi}(x)}S_\tau,\qquad
   {\mit\Gamma}_\tau\frac{\overleftarrow{\delta}}{\delta\Psi(x)}
   =S_\tau\frac{\overleftarrow{\delta}}{\delta\psi(x)}.
\label{eq:(4.5)}
\end{equation}

At this stage, we introduce the following extremely useful ``$-1$ variables''.
These are local functionals of the original field variables with the following
functional dependences:
\begin{align}
   \Psi_{-1}&=\Psi_{-1}[\Psi,L],&
   \Bar{\Psi}_{-1}&=\Bar{\Psi}_{-1}[\Bar{\Psi},L],
\notag\\
   L_{-1,\mu}^a&=L_{-1,\mu}^a[L],&
   \chi_{-1,L}^a&=\chi_{-1,L}^a[\chi_L,L].
\label{eq:(4.6)}
\end{align}
These variables possess the following simple scaling properties:
\begin{subequations}
\label{eq:(4.7)}
\begin{align}
   \left(\frac{D-1}{2}+x\cdot\partial\right)
   \Psi_{-1}(x)
   &=\int d^Dy\,
   \Psi_{-1}(x)\frac{\overleftarrow{\delta}}{\delta\Psi(y)}
   \left(
   \frac{D-1}{2}+y\cdot\partial+2\Delta\right)\Psi(y)
\notag\\
   &\qquad{}
   +\int d^Dy\,
   (1+y\cdot\partial+2\Delta)L_\mu^a(y)\cdot
   \frac{\delta}{\delta L_\mu^a(y)}\,\Psi_{-1}(x),
\\
   \left(\frac{D-1}{2}+x\cdot\partial\right)\Bar{\Psi}_{-1}(x)
   &=\int d^Dy\,
   \Bar{\Psi}(y)
   \left(
   \frac{D-1}{2}+\overleftarrow{\partial}\cdot y+2\overleftarrow{\Delta}
   \right)\frac{\overrightarrow{\delta}}{\delta\Bar{\Psi}(y)}\,
   \Bar{\Psi}_{-1}(x)
\notag\\
   &\qquad{}
   +\int d^Dy\,
   (1+y\cdot\partial+2\Delta)L_\mu^a(y)\cdot
   \frac{\delta}{\delta L_\mu^a(y)}\,\Bar{\Psi}_{-1}(x),
\\
   (1+x\cdot\partial)L_{-1,\mu}^a(x)
   &=\int d^Dy\,
   (1+y\cdot\partial+2\Delta)L_\nu^b(y)\cdot\frac{\delta}{\delta L_\nu^b(y)}\,
   L_{-1,\mu}^a(x),
\end{align}
and
\begin{align}
   x\cdot\partial\chi_{-1,L}^a(x)
   &=\int d^Dy\,
   (y\cdot\partial+2\Delta)\chi_L^b(y)\cdot
   \frac{\delta}{\delta\chi_L^b(y)}\,
   \chi_{-1,L}^a(x)
\notag\\
   &\qquad{}
   +\int d^Dy\,
   (1+y\cdot\partial+2\Delta)L_\mu^b(y)\cdot
   \frac{\delta}{\delta L_\mu^b(y)}\,
   \chi_{-1,L}^a(x).
\end{align}
\end{subequations}
As we explain in~Appendix~\ref{sec:A}, these variables can be obtained by
solving diffusion equations backward in the diffusion time to~$t=-1$,
corresponding to the logarithmic scale~$\tau=-\infty$. The above relations
allow us to rewrite the GFERG equation for the 1PI action in a simple form as
\begin{align}
   \frac{\partial}{\partial\tau}{\mit\Gamma}_\tau
   &=-\int d^Dx\,
   {\mit\Gamma}_\tau
   \frac{\overleftarrow{\delta}}{\delta\Psi_{-1}(x)}
   \left(\frac{D-1}{2}+x\cdot\partial\right)
   \Psi_{-1}(x)
\notag\\
   &\qquad{}
   -\int d^Dx\,
   \Bar{\Psi}_{-1}(x)
   \left(\frac{D-1}{2}
   +\overleftarrow{\partial}\cdot x\right)
   \frac{\overrightarrow{\delta}}{\delta\Bar{\Psi}_{-1}(x)}{\mit\Gamma}_\tau
\notag\\
   &\qquad{}
   -\int d^Dx\,
   \left(1+x\cdot\partial\right)L_{-1,\mu}^a(x)\cdot
   \frac{\delta}{\delta L_{-1,\mu}^a(x)}
   {\mit\Gamma}_\tau
\notag\\
   &\qquad{}
   -\int d^Dx\,\tr\left[
   (1-4\widetilde{\Delta})
   \Psi(x)\cdot
   \frac{\overleftarrow{\delta}}{\delta\psi(x)}\right],
\label{eq:(4.8)}
\end{align}
where
\begin{equation}
   4\widetilde{\Delta}
   \equiv2\left\{
   2\partial^2+\left[\partial_\mu L_\mu(t,x)\right]
   +2L_\mu(t,x)\partial_\mu
   +L_\mu(t,x)L_\mu(t,x)
   +\alpha_0\gamma_5
   \left[\partial_\mu L_\mu(t,x)\right]
   \right\}
\end{equation}
and $L_\mu(t,x)\equiv L_\mu^a(t,x)T^a$. In this expression, we have used
Eq.~\eqref{eq:(4.3)} to simplify the last term, for which
$\Psi(x)\overleftarrow{\delta}/\delta\psi(y)$ is given as the inverse of
\begin{equation}
   \psi(x)\frac{\overleftarrow{\delta}}{\delta\Psi(y)}
   =\delta(x-y)-i\frac{\delta}{\delta\Bar{\Psi}(x)}
   {\mit\Gamma}_\tau\frac{\overleftarrow{\delta}}{\delta\Psi(y)}.
\label{eq:(4.9)}
\end{equation}
For the Wilson action~$S_\tau$ at most bi-linear in the fermion fields, the
corresponding 1PI action~${\mit\Gamma}_\tau$ is also at most bi-linear in the
fermion fields, and thus $\Psi(x)\overleftarrow{\delta}/\delta\psi(y)$ depends
only on the external gauge field.

In Appendix~\ref{sec:A}, we also show that the BRST
transformation\footnote{Here, we do not claim that this transformation is
generated by the BRST transformation on the original field variables $\psi$
and~$\Bar{\psi}$.}
\begin{align}
   \delta\Psi(x)&=-\chi_L^a(x)T^aP_L\Psi(x),&
   \delta\Bar{\Psi}(x)&=\chi_L^a(x)\Bar{\Psi}(x)P_RT^a,
\notag\\
   \delta L_\mu^a(x)
   &=\partial_\mu\chi_L^a(x)+f^{abc}L_\mu^b(x)\chi_L^c(x),&
   \delta\chi_L^a(x)
   &=-\frac{1}{2}f^{abc}\chi_L^b(x)\chi_L^c(x),
\label{eq:(4.10)}
\end{align}
is inherited by the corresponding $-1$ variables as
\begin{align}
   \delta\Psi_{-1}(x)&=-\chi_{-1,L}^a(x)T^aP_L\Psi_{-1}(x),&
   \delta\Bar{\Psi}_{-1}(x)&=\chi_{-1,L}^a(x)\Bar{\Psi}_{-1}(x)P_RT^a,
\notag\\
   \delta L_{-1,\mu}^a(x)
   &=\partial_\mu\chi_{-1,L}^a(x)+f^{abc}L_{-1,\mu}^b(x)\chi_{-1,L}^c(x),&
   \delta\chi_{-1,L}^a(x)
   &=-\frac{1}{2}f^{abc}\chi_{-1,L}^b(x)\chi_{-1,L}^c(x).
\label{eq:(4.11)}
\end{align}
Using the above relations, the anomaly~\eqref{eq:(3.4)} is written as
\begin{align}
   \mathcal{Q}_{L\tau}
   &=\int d^Dx\,
   \Biggl\{
   \chi_{-1,L}^a(x)
   \left[
   -{\mit\Gamma}_\tau
   \frac{\overleftarrow{\delta}}{\delta\Psi_{-1}(x)}T^aP_L\Psi_{-1}(x)
   +\Bar{\Psi}_{-1}(x)P_RT^a
   \frac{\overrightarrow{\delta}}{\delta\Bar{\Psi}_{-1}(x)}
   {\mit\Gamma}_\tau
   \right]
\notag\\
   &\qquad\qquad\qquad{}
   +\left[
   \partial_\mu\chi_{-1,L}^a(x)+f^{abc}L_{-1,\mu}^b(x)\chi_{-1,L}^c(x)
   \right]
   \frac{\delta}{\delta L_{-1,\mu}^a(x)}
   {\mit\Gamma}_\tau
\notag\\
   &\qquad\qquad\qquad{}
   -\chi_L^a(x)\tr
   \left[
   T^a\gamma_5\Psi(x)
   \frac{\overleftarrow{\delta}}{\delta\psi(x)}
   \right]
   \Biggr\}.
\label{eq:(4.12)}
\end{align}

Now, let us split the 1PI action into two parts:
\begin{equation}
   {\mit\Gamma}_\tau[\Psi,\Bar{\Psi},L]
   ={\mit\Gamma}_{f,\tau}[\Psi,\Bar{\Psi},L]
   +{\mit\Gamma}_{g,\tau}[L].
\label{eq:(4.13)}
\end{equation}
The part bi-linear in the fermion fields, ${\mit\Gamma}_{f,\tau}$, can be taken
manifestly invariant under the BRST transformation~\eqref{eq:(4.11)}. For
instance,
\begin{equation}
   {\mit\Gamma}_{f,\tau}[\Psi,\Bar{\Psi},L]
   =i\int d^Dx\,
   \Bar{\Psi}_{-1}(x)\gamma_\mu
   \left[
   \partial_\mu+L_{-1,\mu}^a(x)T^aP_L
   \right]
   \Psi_{-1}(x)
\label{eq:(4.14)}
\end{equation}
is a choice that solves the bi-linear part of the GFERG
equation~\eqref{eq:(4.8)}.\footnote{Note that Eq.~\eqref{eq:(4.14)} is
independent of~$\tau$. We may add terms that become irrelevant
as~$\tau\to+\infty$, such as $e^{-\tau}\int d^Dx\,%
\Bar{\Psi}_{-1}(x)F_{-1,L\mu\nu}^a(x)T^a[\gamma_\mu,\gamma_\nu]\Psi_{-1}(x)$, where
$F_{-1,L\mu\nu}^a$ is the field strength of~$L_{-1,\mu}^a$. Relevant terms such as
$e^\tau\int d^Dx\,\Bar{\Psi}_{-1}(x)\Psi_{-1}(x)$ are not invariant under the
chiral transformation~\eqref{eq:(4.11)}, and are excluded.} With such a choice,
the expression of the chiral anomaly~\eqref{eq:(4.12)} reduces to
\begin{align}
   \mathcal{Q}_{L\tau}
   &=\int d^Dx\,\Biggl\{
   \left[\partial_\mu\chi_{-1,L}^a(x)+f^{abc}L_{-1,\mu}^b(x)\chi_{-1,L}^c(x)\right]
   \frac{\delta}{\delta L_{-1,\mu}^a(x)}
   {\mit\Gamma}_{g,\tau}
\notag\\
   &\qquad\qquad\qquad{}
   -\chi_L^a(x)\tr
   \left[
   T^a\gamma_5\Psi(x)
   \frac{\overleftarrow{\delta}}{\delta\psi(x)}
   \right]
   \Biggr\},
\label{eq:(4.15)}
\end{align}
which is completely independent of the fermion fields. Then, the WZ
condition~\eqref{eq:(3.6)} simplifies to
\begin{align}
   &\int d^Dx\,\Biggl\{
   \left[\partial_\mu\chi_{-1,L}^a(x)
   +f^{abc}L_{-1,\mu}^b(x)\chi_{-1,L}^c(x)\right]
   \frac{\delta}{\delta L_{-1,\mu}^a(x)}
\notag\\
   &\qquad\qquad\qquad{}
   -\frac{1}{2}f^{abc}\chi_{-1,L}^b(x)\chi_{-1,L}^c(x)
   \frac{\delta}{\delta\chi_{-1,L}^a(x)}
   \Biggr\}
   \mathcal{Q}_{L\tau}=0,
\label{eq:(4.16)}
\end{align}
and the GFERG equation~\eqref{eq:(3.18)} reduces to
\begin{align}
   \frac{\partial}{\partial\tau}\mathcal{Q}_{L\tau}
   &=-\int d^Dx\,
   \left(1+x\cdot\partial\right)L_{-1,\mu}^a(x)\cdot
   \frac{\delta}{\delta L_{-1,\mu}^a(x)}
   \mathcal{Q}_{L\tau}
\notag\\
   &\qquad{}
   -\int d^Dx\,
   x\cdot\partial\chi_{-1,L}^a(x)\cdot
   \frac{\delta}{\delta\chi_{-1,L}^a(x)}
   \mathcal{Q}_{L\tau}.
\label{eq:(4.17)}
\end{align}

Now, we can repeat the above argument by introducing another external gauge
field~$R_\mu^a(x)$, which couples to the right-handed fermion current. All the
considerations go through with trivial changes: the 1PI action now depends also
on~$R_\mu^a$, and the associated anomaly~$\mathcal{Q}_R$ is proportional to the
ghost~$\chi_R^a$ for the $\mathrm{SU}(N)_R$ transformation, etc. To summarize,
we obtain
\begin{equation}
   {\mit\Gamma}_{f,\tau}[\Psi,\Bar{\Psi},L,R]
   =i\int d^Dx\,
   \Bar{\Psi}_{-1}(x)\gamma_\mu
   \left\{
   \partial_\mu
   +T^a\left[L_{-1,\mu}^a(x)P_L+R_{-1,\mu}^a(x)P_R\right]\right\}\Psi_{-1}(x),
\label{eq:(4.18)}
\end{equation}
the gauge field part of the action satisfies the GFERG equation
\begin{align}
   &\frac{\partial}{\partial\tau}{\mit\Gamma}_{g,\tau}
\notag\\
   &=-\int d^Dx\,
   \left[
   (1+x\cdot\partial)L_{-1,\mu}^a(x)\cdot
   \frac{\delta}{\delta L_{-1,\mu}^a(x)}
   +(1+x\cdot\partial)R_{-1,\mu}^a(x)\cdot
   \frac{\delta}{\delta R_{-1,\mu}^a(x)}
   \right]
   {\mit\Gamma}_{g,\tau}
\notag\\
   &\qquad{}
   -\int d^D x\,
   \tr\left[
   (1-4\widetilde{\Delta})\Psi(x)\cdot
   \frac{\overleftarrow{\delta}}{\delta\psi(x)}
   \right],
\label{eq:(4.19)}
\end{align}
and the anomaly, defined by
\begin{align}
   \mathcal{Q}_\tau
   &\equiv
   \mathcal{Q}_{L\tau}+\mathcal{Q}_{R\tau}
\notag\\
   &=\int d^Dx\,
   \Biggl\{
   \left[
   \partial_\mu\chi_{-1,L}^a(x)
   +f^{abc}L_{-1,\mu}^b(x)\chi_{-1,L}^c(x)\right]
   \frac{\delta{\mit\Gamma}_{g,\tau}}{\delta L_{-1,\mu}^a(x)}
\notag\\
   &\qquad\qquad\qquad{}
   +\left[
   \partial_\mu\chi_{-1,R}^a(x)
   +f^{abc}R_{-1,\mu}^b(x)\chi_{-1,R}^c(x)\right]
   \frac{\delta{\mit\Gamma}_{g,\tau}}{\delta R_{-1,\mu}^a(x)}
\notag\\
   &\qquad\qquad\qquad{}
   +\left[\chi_R^a(x)-\chi_L^a(x)\right]
   \tr\left[
   T^a\gamma_5\Psi(x)\frac{\overleftarrow{\delta}}{\delta\psi(x)}
   \right]\Biggr\},
\label{eq:(4.20)}
\end{align}
satisfies both the GFERG equation
\begin{align}
   &\frac{\partial}{\partial\tau}\mathcal{Q}_\tau
\notag\\
   &=-\int d^Dx\,
   \Biggl[
   (1+x\cdot\partial)L_{-1,\mu}^a(x)\cdot
   \frac{\delta}{\delta L_{-1,\mu}^a(x)}
   +(1+x\cdot\partial)R_{-1,\mu}^a(x)\cdot
   \frac{\delta}{\delta R_{-1,\mu}^a(x)}
\notag\\
   &\qquad\qquad\qquad{}
   +x\cdot\partial\chi_{-1,L}^a(x)\cdot\frac{\delta}{\delta\chi_{-1,L}^a(x)}
   +x\cdot\partial\chi_{-1,R}^a(x)\cdot\frac{\delta}{\delta\chi_{-1,R}^a (x)}
   \Biggr]
   \mathcal{Q}_\tau,
\label{eq:(4.21)}
\end{align}
and the WZ consistency condition
\begin{align}
   &\int d^Dx\,
   \Biggl\{
   \left[\partial_\mu\chi_{-1,L}^a(x)
   +f^{abc}L_{-1,\mu}^b(x)\chi_{-1,L}^c(x)\right]
   \frac{\delta}{\delta L_{-1,\mu}^a(x)}
\notag\\
   &\qquad\qquad\qquad{}
   -\frac{1}{2}f^{abc}\chi_{-1,L}^b(x)\chi_{-1,L}^c(x)
   \frac{\delta}{\delta\chi_{-1,L}^a(x)}
\notag\\
   &\qquad\qquad{}
   +\left[\partial_\mu\chi_{-1,R}^a(x)
   +f^{abc}R_{-1,\mu}^b(x)\chi_{-1,R}^c(x)\right]
   \frac{\delta}{\delta R_{-1,\mu}^a(x)}
\notag\\
   &\qquad\qquad\qquad{}
   -\frac{1}{2}f^{abc}\chi_{-1,R}^b(x)\chi_{-1,R}^c(x)
   \frac{\delta}{\delta\chi_{-1,R}^a(x)}
   \Biggr\}
   \mathcal{Q}_\tau=0.
\label{eq:(4.22)}
\end{align}
In the following, we would like to show that $\mathcal{Q}_\tau$ is a scale
invariant functional of~$L_{-1,\mu}^a$ and~$R_{-1,\mu}^a$ with a
$\tau$-independent overall constant.

To begin with, we note that ${\mit\Gamma}_{f,\tau}$~\eqref{eq:(4.18)} has no
explicit $\tau$~dependence. Hence, neither does
$\psi(x)\overleftarrow{\delta}/\delta\Psi(y)$ given
by~Eq.~\eqref{eq:(4.9)}. Inverting this we obtain $\tau$-independent
$\Psi(x)\overleftarrow{\delta}/\delta\psi(y)$. We can then expand
\begin{equation}
   -\int d^Dx\,\tr
   \left[
   (1-4\widetilde{\Delta})
   \Psi(x)\cdot\frac{\overleftarrow{\delta}}{\delta\psi(x)}
   \right]
   =\sum_dA_d[L,R],
\label{eq:(4.23)}
\end{equation}
where $A_d[L,R]$ is a $\tau$-independent functional of~$L_{-1\mu}^a$
and~$R_{-1\mu}^a$ with scaling dimension~$d$, satisfying
\begin{align}
   &\int d^Dx\,
   \left[
   (1+x\cdot\partial)L_{-1,\mu}^a(x)\frac{\delta}{\delta L_{-1,\mu}^a(x)}
   +(1+x\cdot\partial)R_{-1,\mu}^a(x)\frac{\delta}{\delta R_{-1,\mu}^a(x)}
   \right]A_d[L,R]
\notag\\
   &=dA_d[L,R].
\label{eq:(4.24)}
\end{align}
Then, Eq.~\eqref{eq:(4.19)} implies
\begin{equation}
   {\mit\Gamma}_{g,\tau}[L,R]
   =\sum_{d\neq0}\frac{1}{d}A_d[L,R]
   +\tau A_0[L,R]+\sum_de^{-d\tau}A_d'[L,R],
\label{eq:(4.25)}
\end{equation}
where the last sum is a homogeneous solution to~Eq.~\eqref{eq:(4.19)}
corresponding to local counterterms. We can choose to remove $A_d'[L,R]$
for~$d\neq0$ to obtain
\begin{equation}
   {\mit\Gamma}_{g,\tau}[L,R]
   =\sum_{d\neq0}\frac{1}{d}A_d[L,R]+\tau A_0[L,R]+A_0'[L,R].
\label{eq:(4.26)}
\end{equation}
We keep $A_0'[L,R]$ to simplify the form of the anomaly later.

Thus, the $\tau$ dependent part, $A_0[L,R]$, of~${\mit\Gamma}_{g,\tau}$ has
scaling dimension~$0$. Differentiating Eq.~\eqref{eq:(4.20)} with respect
to~$\tau$ and using Eq.~\eqref{eq:(4.26)}, we obtain
\begin{align}
   \frac{\partial}{\partial\tau}\mathcal{Q}_\tau
   &=\int d^Dx\,
   \Biggl\{
   \left[\partial_\mu\chi_{-1,L}^a(x)
   +f^{abc}L_{-1,\mu}^b(x)\chi_{-1,L}^c(x)
   \right]\frac{\delta}{\delta L_{-1,\mu}^a(x)}
\notag\\
   &\qquad\qquad\qquad{}
   +\left[
   \partial_\mu\chi_{-1,R}^a(x)+f^{abc}R_{-1,\mu}^b(x)\chi_{-1,R}^c(x)
   \right]\frac{\delta}{\delta R_{-1,\mu}^a(x)}
   \Biggr\}
   A_0[L,R]
\notag\\
   &\equiv B_0[L,R,\chi_L,\chi_R],
\label{eq:(4.27)}
\end{align}
which is a $\tau$-independent functional with scaling dimension $0$. This
implies we can expand
\begin{equation}
   \mathcal{Q}_\tau=\tau B_0[L,R,\chi_L,\chi_R]+\sum_d C_d[L,R,\chi_L,\chi_R],
\label{eq:(4.28)}
\end{equation}
where $C_d$ is a functional with scaling dimension~$d$. Substituting this
result into~Eq.~\eqref{eq:(4.21)}, we obtain
\begin{equation}
   B_0=-\sum_ddC_d,
\label{eq:(4.29)}
\end{equation}
implying
\begin{equation}
   B_0=C_{d\neq0}=0,
\label{eq:(4.30)}
\end{equation}
and
\begin{equation}
   \mathcal{Q}_\tau=C_0[L,R,\chi_L,\chi_R].
\label{eq:(4.31)}
\end{equation}
Thus, the anomaly is a $\tau$-independent functional with scaling
dimension~$0$. This is what we call the scale invariance of the anomaly. By
definition, the anomaly is linear in $\chi_{-1 L}^a$ and~$\chi_{-1 R}^a$. Since
the GFERG keeps the vector gauge invariance manifestly, we also know that the
anomaly vanishes if $L_\mu^a=R_\mu^a$ and~$\chi_L=\chi_R$:
\begin{equation}
   C_0[V,V,\chi,\chi]=0.
\label{eq:(4.32)}
\end{equation}

We now recall that Eq.~\eqref{eq:(4.19)} leaves ${\mit\Gamma}_\tau$ ambiguous
by the addition of a scale invariant functional~$A_0'[L,R]$
in~Eq.~\eqref{eq:(4.26)}. Since the anomaly satisfies the WZ
condition~\eqref{eq:(4.22)}, by choosing $A_0'[L,R]$ appropriately, we can make
${\mit\Gamma}_{g,\tau}$ invariant under the vector transformations in the
presence of both left and right gauge fields:
\begin{equation}
   C_0[L,R,\chi,\chi]=0.
\label{eq:(4.33)}
\end{equation}
Now, the anomaly arises only in the axial transformations parametrized by
\begin{equation}
   \chi_A^a(x)\equiv\frac{1}{2}\left[\chi_R^a(x)-\chi_L^a(x)\right].
\label{eq:(4.34)}
\end{equation}
We thus obtain the anomaly $\mathcal{Q}_\tau$ as a scale invariant functional,
denoted by~$\mathcal{Q}_A$, of~$L_{-1,\mu}^a$, $R_{-1,\mu}^a$,
and~$\chi_{-1,A}^a$. In~$D=4$, such a combination is well-known, and is given by
the Bardeen form of the gauge anomaly~\cite{Bardeen:1969md}, written in terms
of the $-1$~variables as\footnote{In what follows, the convention
$X\equiv X^aT^a$ is to be understood.}
\begin{align}
   \mathcal{Q}_A
   &=\mathcal{A}
   \int d^4x\,\varepsilon_{\mu\nu\rho\sigma}
   \tr\biggl\{\chi_{-1,A}\biggl[
   \frac{1}{4}F_{-1,V\mu\nu}F_{-1,V\rho\sigma}
   +\frac{1}{12}F_{-1,A\mu\nu}F_{-1,A\rho\sigma}
\notag\\
   &\qquad\qquad{}
   -\frac{2}{3}
   \left(
   F_{-1,V\mu\nu}A_{-1,\rho}A_{-1,\sigma}
   +A_{-1,\mu}F_{-1,V\nu\rho}A_{-1,\sigma}
   +A_{-1,\mu}A_{-1,\nu}F_{-1,V\rho\sigma}
   \right)
\notag\\
   &\qquad\qquad{}
   +\frac{8}{3}A_{-1,\mu}A_{-1,\nu}A_{-1,\rho}A_{-1,\sigma}
   \biggr]\biggr\},
\label{eq:(4.35)}
\end{align}
where $\mathcal{A}$ is a $\tau$-independent real constant, and
\begin{subequations}
\label{eq:(4.36)}
\begin{align}
   \chi_{-1,A}&\equiv\frac{1}{2}(\chi_{-1,R}-\chi_{-1,L}),&&
\\
   F_{-1,V\mu\nu}&\equiv
   \frac{1}{2}(F_{-1,R\mu\nu}+F_{-1,L\mu\nu}),&
   F_{-1,A\mu\nu}&\equiv
   \frac{1}{2}(F_{-1,R\mu\nu}-F_{-1,L\mu\nu}),
\\
   V_{-1,\mu}&\equiv\frac{1}{2}(R_{-1,\mu}+L_{-1,\mu}),&
   A_{-1,\mu}&\equiv\frac{1}{2}(R_{-1,\mu}-L_{-1,\mu}).
\end{align}
\end{subequations}
$F_{-1,R\mu\nu}$ and~$F_{-1,L\mu\nu}$ are the field strengths of~$R_{-1,\mu}$
and~$L_{-1,\mu}$, respectively.

\subsection{Computation of~$\mathcal{A}$}
The coefficient~$\mathcal{A}$ in~Eq.~\eqref{eq:(4.35)} can be determined by
choosing a particular gauge field configuration, $R_\mu=L_\mu=V_\mu$. The
anomaly is then simplified to
\begin{equation}
   \mathcal{Q}_A
   =\frac{1}{4}\mathcal{A}
   \int d^4 x\,
   \varepsilon_{\mu\nu\rho\sigma}
   \tr\left(\chi_{-1,A}F_{-1,V\mu\nu}F_{-1,V\rho\sigma}
   \right).
\label{eq:(4.37)}
\end{equation}
We may further set the $-1$~variables to the lowest order in gauge fields; for
the choice~$\alpha_0=1$,\footnote{The parameter $\alpha_0$ affects only the
dependence of the $-1$~variables on $L_\mu^a$ and~$R_\mu^a$, but not
$\mathcal{A}$.} we have $V_{-1,\mu}(x)=e^{-\partial^2}V_\mu(x)+\dotsb$
and~$\chi_{-1,A}(x)=e^{-\partial^2}\chi_A(x)+\dotsb$. Hence, in momentum space, we
find
\begin{equation}
   \mathcal{Q}_A
   =-\mathcal{A}
   \int_{k,l}\,\varepsilon_{\mu\nu\rho\sigma}
   \tr\left[
   e^{(k+l)^2}\chi_A(-k-l)e^{k^2}k_\mu V_\nu(k)
   e^{l^2}l_\rho V_\sigma(l)
   \right]+\dotsb,
\label{eq:(4.38)}
\end{equation}
where the terms higher order in gauge fields are suppressed. On the other hand,
Eq.~\eqref{eq:(4.20)} gives
\begin{align}
   \mathcal{Q}_A
   &=\int d^Dx\,
   \Biggl\{
   -\chi_{-1,A}^a(x)\left[\delta^{ac}\partial_\mu+f^{abc}V_{-1,\mu}^b(x)\right]
   \frac{\delta}{\delta A_{-1,\mu}^c(x)}{\mit\Gamma}_{g,\tau}
\notag\\
   &\qquad\qquad\qquad{}
   +2\tr
   \left[
   \chi_A(x)\gamma_5\Psi(x)
   \frac{\overleftarrow{\delta}}{\delta\psi(x)}
   \right]
   \Biggr\}.
\label{eq:(4.39)}
\end{align}
The first term, being the covariant derivative of a local
term~$[\delta/\delta A_{-1,\mu}^c(x)]{\mit\Gamma}_{g,\tau}$, gives no
contribution to the lowest order in~$(k+l)^2$ and can be neglected for the
determination of~$\mathcal{A}$ in~Eq.~\eqref{eq:(4.38)}.

For the last term of~\eqref{eq:(4.39)},
$\Psi(x)\overleftarrow{\delta}/\delta\psi(y)$ is obtained by inverting
\begin{align}
   &\psi(x)\frac{\overleftarrow{\delta}}{\delta\Psi(y)}
\notag\\
   &=\left[\Psi(x)-i\frac{\overrightarrow{\delta}}{\delta\Bar{\Psi}(x)}
   {\mit\Gamma}_{f,\tau}\right]
   \frac{\overleftarrow{\delta}}{\delta\Psi(y)}
\notag\\
   &=\delta(x-y)+\frac{\delta}{\delta\Bar{\Psi}(x)}
   \int d^Dz\,
   \Bar{\Psi}_{-1}(z)\gamma_\mu
   \left[
   \partial_\mu+V_{-1,\mu}(z)
   \right]
   \Psi_{-1}(z)
   \frac{\overleftarrow{\delta}}{\delta\Psi(y)}
\notag\\
   &=\int_p\,e^{ip(x-y)}e^{p^2}
   \left(e^{-2p^2}+i\Slash{p}\right)e^{p^2}
\notag\\
   &\qquad{}
   +\int_{p,k}\,
   e^{i(p+k)x-ipy}
   e^{(p+k)^2}\Bar{\mathcal{V}}_\mu(-p-k,k,p)e^{k^2}V_\mu(k)e^{p^2}
   +O(V^2),
\label{eq:(4.40)}
\end{align}
where we have used Eq.~\eqref{eq:(4.18)}
with~$R_{-1,\mu}^a=L_{-1,\mu}^a=V_{-1,\mu}^a$. The vertex function
$\Bar{\mathcal{V}}_\mu$ is obtained by solving the diffusion
equations~\eqref{eq:(A1)} in~Appendix~\ref{sec:A} to~first order in~$V$. The
result is
\begin{align}
   &\Bar{\mathcal{V}}_\mu(-p-k,k,p)
   =\Bar{\mathcal{V}}_\mu(-p,-k,p+k)
\notag\\
   &=\gamma_\mu
   +2p_\mu(\Slash{p}+\Slash{k})F(2p\cdot k)
   +2(p+k)_\mu\Slash{p}F(-2(p+k)\cdot k),
\label{eq:(4.41)}
\end{align}
where
\begin{equation}
   F(x)\equiv\frac{e^x-1}{x}.
\label{eq:(4.42)}
\end{equation}
($\Bar{\mathcal{V}}_\mu$ can also be found~in~Ref.~\cite{Sonoda:2022fmk}.)
Since the diffusion equations~\eqref{eq:(A1)} do not contain any Dirac
matrices, the higher order $O(V^2)$ terms in~Eq.~\eqref{eq:(4.40)} are also
linear in~$\gamma_\mu$. But the trace in the last term
of~Eq.~\eqref{eq:(4.39)} requires at least four Dirac matrices, and the
$O(V^2)$~term in~Eq.~\eqref{eq:(4.40)} gives no contribution. We then obtain
\begin{align}
   \mathcal{Q}_A
   &=2\int_{k,l}\,
   \tr\left[\chi_{-1,A}(-k-l)e^{k^2}V_\mu(k)e^{l^2}V_\nu(l)\right]
   \int_p\,e^{-p^2}e^{-(p+k+l)^2}
\notag\\
   &\qquad{}
   \times
   \tr\biggl\{
   \gamma_5
   \left[e^{-2(p+k+l)^2}+i(\Slash{p}+\Slash{k}+\Slash{l})\right]^{-1}
   \Bar{\mathcal{V}}_\mu(-p-k-l,k,p+l)
\notag\\
   &\qquad\qquad\qquad{}
   \times
   \left[e^{-2(p+l)^2}+i(\Slash{p}+\Slash{l})\right]^{-1}
   \Bar{\mathcal{V}}_\nu(-p-l,l,p)
   \left(e^{-2p^2}+i\Slash{p}\right)^{-1}
   \biggr\}+\dotsb.
\label{eq:(4.43)}
\end{align}
To the first order in $k$ and~$l$, the integral has been calculated
in~Ref.~\cite{Miyakawa:2022qbz} as\footnote{Our convention is
$\gamma_5=\gamma_1\gamma_2\gamma_3\gamma_4$, and~$\varepsilon_{1234}=%
(1/4)\tr\gamma_5\gamma_1\gamma_2\gamma_3\gamma_4=1$.}
\begin{align}
   &\int_p\,e^{-(p-l)^2}e^{-(p+k)^2}
   \tr\biggl\{
   \gamma_5
   \left[e^{-2(p+k)^2}+i(\Slash{p}+\Slash{k})\right]^{-1}
   \Bar{\mathcal{V}}_\mu(-p-k,k,p)
\notag\\
   &\qquad\qquad\qquad\qquad\qquad{}
   \times
   \left[e^{-2p^2}+i\Slash{p}\right]^{-1}
   \Bar{\mathcal{V}}_\nu(-p+l,-l,p)
   \left(e^{-2(p-l)^2}+i(\Slash{p}-\Slash{l})\right)^{-1}
   \biggr\}
\notag\\
   &\stackrel{k,l\to0}{\longrightarrow}
   \tr\gamma_5\gamma_\mu\gamma_\nu\Slash{k}\Slash{l}
   \int_p\,\frac{e^{-4p^2}}{(e^{-4p^2}+p^2)^3}(1+4p^2)
   =\frac{2}{(4\pi)^2}\varepsilon_{\mu\nu\alpha\beta}k_\alpha l_\beta,
\label{eq:(4.44)}
\end{align}
where the integral is evaluated as
\begin{equation}
   \int_p\,\frac{e^{-4p^2}}{\left(e^{-4p^2}+p^2\right)^3}(1+4p^2)
   =\frac{1}{(4\pi)^2}\int_0^\infty d\xi\,\left(-\frac{1}{2}\right)
   \frac{d}{d\xi}
   \frac{1+2\xi e^{4\xi}}{(1+\xi e^{4\xi})^2}
   =\frac{1}{(4\pi)^2}\frac{1}{2}.
\label{eq:(4.45)}
\end{equation}
Comparing the result with~Eq.~\eqref{eq:(4.38)}, we obtain
\begin{equation}
   \mathcal{A}=\frac{1}{4\pi^2}.
\label{eq:(4.46)}
\end{equation}

\section{Comment on the inclusion of QCD}
\label{sec:5}
So far, we have considered the massless free fermions coupled only to external
gauge fields. Before concluding the paper we would like to comment on how our
argument and conclusion should (or should not) be changed when we couple the
fermions to \emph{dynamical\/} gauge fields such as the gluons in
QCD~\cite{Sonoda:2020vut,Miyakawa:2021hcx}. First of all, in defining the
Wilson action~\eqref{eq:(2.1)}, we should change the canonical scaling factor
$[(D-1)/2](\tau-\tau_0)$ by including the anomalous dimension
$\int_{\tau_0}^\tau d\tau'\,[(D-1)/2+\gamma_{\tau'}]$. We also get the Gaussian
integration over the dynamical gauge fields $v_\mu^A (x)$. Correspondingly, the
unscrambler~\eqref{eq:(2.2)} must have the additional
factor~$\exp\{-(1/2)\int d^Dx\,\delta^2/[\delta v_\mu^A(x)\delta
v_\mu^A(x)]\}$ for the dynamical gauge fields. The diffusion equation for the
fermi fields depends also on the dynamical gauge fields, which satisfy their
own diffusion equation.

The external gauge transformation remains generated by~Eq.~\eqref{eq:(3.1)} and
commutes with the diffusion equations. The chiral anomaly~$\mathcal{Q}_{L\tau}$,
still given by~Eq.~\eqref{eq:(3.4)}, remains a composite operator with scaling
dimension~$0$. The GFERG equation for~$\mathcal{Q}_{L\tau}$, however, is no
longer given by~Eq.~\eqref{eq:(3.18)}; the scaling dimension of the fermions is
changed from~$(D-1)/2$ to~$(D-1)/2+\gamma_\tau$, and we get an additional term
of functional differentiation with respect to the dynamical gauge fields. We
can introduce the $-1$~variables for the dynamical gauge fields by repeating
the argument given in~Appendix~\ref{sec:A}, but a simple construction such
as~Eq.~\eqref{eq:(4.14)} does not solve the GFERG equation anymore. Though we
believe Eq.~\eqref{eq:(4.15)} is still valid, its derivation does not go
through unchanged.

However, \emph{if we assume\/} that $\mathcal{Q}_{\tau}$ depends only on the
external gauge fields~$L_\mu^a$ and $R_\mu^a$, then $\mathcal{Q}_{\tau}$ again
satisfies Eqs.~\eqref{eq:(4.20)}, \eqref{eq:(4.21)}, and~\eqref{eq:(4.22)}. We
can now repeat the above argument and infer that Eq.~\eqref{eq:(4.35)} is the
most general form of the anomaly in the presence of both left and right gauge
fields. Then Eq.~\eqref{eq:(4.21)} implies that the anomaly
coefficient~$\mathcal{A}$ has no $\tau$ dependence. This implies further that
$\mathcal{A}$ cannot depend on the $\tau$-dependent gauge coupling
constant~$g_\tau$ to the dynamical gauge fields. We can thus conclude that the
chiral anomaly is not affected by the inclusion of the dynamical gauge fields.
This is the statement of the Adler--Bardeen theorem~\cite{Zee:1972zt}.

\section{Conclusion}
\label{sec:6}
In this paper we have successfully tested the GFERG formalism by deriving the
chiral anomaly for the free massless fermions under arbitrary background left
and right gauge fields. We have formulated the anomaly as a composite operator
that generates the chiral gauge (BRST) transformations, and have shown that its
scaling dimension vanishes, i.e., the chiral anomaly is scale invariant. We
have also shown that the standard form of the anomaly is obtained in terms of
the $-1$~variables which are functionals of the diffusing external gauge
fields. As in the ERG formalism, the chiral anomaly is expressed by a finite
loop integral that we can calculate without taking a short distance limit.

Once we formulate QCD in the GFERG formalism, we intend to extend the results
of this paper as sketched in~Sect.~\ref{sec:5}.

\section*{Acknowledgments}
Discussions during the YITP/RIKEN iTHEMS workshop ``Lattice and continuum field
theories 2022'' (YITP-W-22-02) were quite useful in constructing this work.
This work was partially supported by Japan Society for the Promotion of Science
(JSPS) Grant-in-Aid for Scientific Research Grant Numbers JP20H01903
and~JP23K03418.
The work of Y.M. was supported by a Kyushu University Innovator Fellowship in
Quantum Science.

\appendix

\section{$-1$ variables}
\label{sec:A}
In this Appendix, we give a precise definition of the ``$-1$ variables''
employed in our construction of the anomaly in the main text and derive their
transformation properties under the scale transformation and the BRST
transformation~\eqref{eq:(4.10)}.

We consider the diffusion equations,
\begin{subequations}
\label{eq:(A1)}
\begin{align}
   \partial_t\Psi(t,x)
   &=\left\{
   \left[\partial_\mu+L_\mu^a(t,x)T^aP_L
   \right]^2
   -\alpha_0\partial_\mu L_\mu^a(t,x)T^aP_L
   \right\}
   \Psi(t,x)
   \equiv\Delta\Psi(t,x),
\\
   \partial_t\Bar{\Psi}(t,x)
   &=\Bar{\Psi}(t,x)
   \left\{
   \left[\overleftarrow{\partial}_\mu
   -P_RL_\mu^a(t,x)T^a
   \right]^2
   +\alpha_0P_R\partial_\mu L_\mu^a(t,x)T^a
   \right\}
   \equiv\Bar{\Psi}(t,x)\overleftarrow{\Delta},
\\
   \partial_tL_\mu^a(t,x)
   &=D_{L\nu}F_{L\nu\mu}^a(t,x)
   +\alpha_0D_{L\mu}\partial_\nu L_\nu^a(t,x)
   \equiv\Delta L_\mu^A(t,x),
\\
   \partial_t\chi_L^a(t,x)
   &=\alpha_0D_{L\mu}\partial_\mu\chi_L^a(t,x)
   \equiv\Delta\chi_L^a(t,x),
\end{align}
\end{subequations}
where
\begin{subequations}
\label{eq:(A2)}
\begin{align}
   F_{L\mu\nu}^a(t,x)
   &\equiv
   \partial_\mu L_\nu^a(t,x)-\partial_\nu L_\mu^a(t,x)
   +f^{abc}L_\mu^b(t,x)L_\nu^c(t,x),
\\
   D_{L\mu}X^a(t,x)
   &\equiv
   \partial_\mu X^a(t,x)
   +f^{abc}L_\mu^b(t,x)X^c(t,x).
\end{align}
\end{subequations}
The first two have the same form as the diffusion equations~\eqref{eq:(2.3)}
for the original field variables $\psi$ and~$\Bar{\psi}$. The initial
conditions for these equations are given by the field variables in the 1PI
formalism:
\begin{align}
   \Psi(0,x)&=\Psi(x),&\Bar{\Psi}(0,x)&=\Bar{\Psi}(x),
\notag\\
   L_\mu^a(0,x)&=L_\mu^a(x),&\chi_L^a(0,x)&=\chi_L^a(x).
\label{eq:(A3)}
\end{align}
The $-1$ variables are simply given by the set of solutions
to~Eqs.~\eqref{eq:(A1)} at~$t=-1$:
\begin{align}
   \Psi_{-1}(x)&\equiv\Psi(-1,x),&
   \Bar{\Psi}_{-1}(x)&\equiv\Bar{\Psi}(-1,x),
\notag\\
   L_{-1,\mu}^a(x)&\equiv L_\mu^a(-1,x),&\chi_{-1,L}^a(x)&\equiv\chi_L^a(-1,x).
\label{eq:(A4)}
\end{align}
This results in the functional dependence of these variables given
by~Eq.~\eqref{eq:(4.6)}.

As is explained in the main text, the BRST transformation~\eqref{eq:(3.9)} is
consistent with the diffusion equations satisfied by $\psi'$, $\Bar{\psi}'$,
$L'$, and~$\chi'_L$. Analogously, these diffusion equations are invariant under
the BRST transformation given by~Eq.~\eqref{eq:(4.10)}. This immediately
implies that the initial values~\eqref{eq:(A3)} and the
solution~\eqref{eq:(A4)} transform in the same way,
implying~Eq.~\eqref{eq:(4.11)}.

Let us derive the scaling properties of the $-1$~variables, given
by~Eqs.~\eqref{eq:(4.7)}. Given a set of solutions $\Psi(t,x)$,
$\Bar{\Psi}(t,x)$, $L_\mu^a(t,x)$, and~$\chi_L^a(t,x)$ to~Eq.~\eqref{eq:(A1)},
we can construct a new set
\begin{subequations}
\label{eq:(A5)}
\begin{align}
   \Psi'(t,x)
   &\equiv e^{[(D-1)/2]\tau}\Psi(e^{2\tau}t,e^\tau x),
\\
   \Bar{\Psi}'(t,x)
   &\equiv e^{[(D-1)/2]\tau}\Bar{\Psi}(e^{2\tau}t,e^\tau x),
\\
   L_\mu^{\prime a}(t,x)
   &\equiv e^\tau L_\mu^a(e^{2\tau}t,e^\tau x),
\\
   \chi^{\prime a}(t,x)
   &\equiv\chi^a(e^{2\tau}s,e^\tau x),
\end{align}
\end{subequations}
by scaling the variables $t\to e^{2\tau}t$, $x\to e^\tau x$ and multiplying the
fermion fields by an appropriate factor~$e^{[(D-1)/2]\tau}$. The primed fields
satisfy Eq.~\eqref{eq:(A1)}. We can then shift the diffusion time to construct
yet another set of solutions:
\begin{subequations}
\label{eq:(A6)}
\begin{align}
   \Psi''(t,x)
   &\equiv\Psi'(t+1-e^{-2\tau},x)
   =e^{[(D-1)/2]\tau}\Psi(e^{2\tau}(t+1)-1,e^\tau x),
\\
   \Bar{\Psi}''(t,x)
   &\equiv\Bar{\Psi}'(t+1-e^{-2\tau},x)
   =e^{[(D-1)/2]\tau}\Bar{\Psi}(e^{2\tau}(t+1)-1,e^\tau x),
\\
   L_\mu^{\prime\prime a}(t,x)
   &\equiv L_\mu^{\prime a}(t+1-e^{-2\tau}, x)
   =e^\tau L_\mu^a(e^{2\tau}(t+1)-1,e^\tau x),
\\
   \chi^{\prime\prime a}(t,x)
   &\equiv\chi^{\prime a}(t+1-e^{-2\tau},x)
   =\chi^{\prime a}(e^{2\tau}(t+1)-1,e^\tau x).
\end{align}
\end{subequations}

Recall that we can regard the double primed fields at~$t=-1$ as functionals of
those at~$t=0$. For an infinitesimal~$\tau$, we find
\begin{align}
   \Psi''(-1,x)
   &=e^{[(D-1)/2]\tau}\Psi(-1,e^\tau x)
\notag\\
   &\simeq\Psi(-1,x)+\tau\left(\frac{D-1}{2}+x\cdot\partial\right)\Psi(-1,x),
\label{eq:(A7)}
\end{align}
and
\begin{align}
   \Psi''(0,x)
   &=e^{[(D-1)/2]\tau}\Psi(e^{2\tau}-1,e^\tau x)
\notag\\
   &\simeq\Psi(0,x)+\tau\left(\frac{D-1}{2}+2\partial_t+x\cdot\partial\right)
   \Psi(0,x)
\notag\\
   &=\Psi(x)+\tau\left(\frac{D-1}{2}+x\cdot\partial+2\Delta\right)\Psi(x).
\label{eq:(A8)}
\end{align}
Similarly, we obtain
\begin{align}
   \Bar{\Psi}''(-1,x)
   &\simeq\Bar{\Psi}(-1,x)
   +\tau\left(\frac{D-1}{2}+x\cdot\partial\right)\Bar{\Psi}(-1,x),
\notag\\
   \Bar{\Psi}''(0,x)
   &\simeq\Bar{\Psi}(x)
   +\tau\Bar{\Psi}(x)
   \left(
   \frac{D-1}{2}+\overleftarrow{\partial}\cdot x+2\overleftarrow{\Delta}
   \right),
\notag\\
   L_\mu^{\prime\prime a}(-1,x)
   &\simeq L_\mu^a(-1,x)
   +\tau(1+x\cdot\partial)L_\mu^a(-1,x),
\notag\\
   L_\mu^{\prime\prime a}(0,x)
   &\simeq L_\mu^a(x)
   +\tau(1+x\cdot\partial+2\Delta)L_\mu^a(x),
\notag\\
   \chi^{\prime\prime a}(-1,x)
   &\simeq\chi^a(-1,x)
   +\tau x\cdot\partial\chi^a(-1,x),
\notag\\
   \chi^{\prime\prime a}(0,x)
   &\simeq\chi^a(x)
   +\tau(x\cdot\partial+2\Delta)\chi^a(x).
\label{eq:(A9)}
\end{align}
Regarding $\Psi_{-1}(x)=\Psi(-1,x)$ as a functional of~$\Psi(x)$
and~$L_\mu^a(x)$, and comparing the coefficients of~$\tau$
in~Eqs.~\eqref{eq:(A7)}, \eqref{eq:(A8)}, and~$L_\mu^{\prime\prime a}(0,x)$
in~Eq.~\eqref{eq:(A9)}, we obtain
\begin{align}
   \left(\frac{D-1}{2}+x\cdot\partial\right)\Psi_{-1}(x)
   &=\int d^Dy\,
   \Biggl[
   \Psi_{-1}(x)\frac{\overleftarrow{\delta}}{\delta\Psi(y)}
   \left(
   \frac{D-1}{2}+y\cdot\partial+2\Delta\right)\Psi(y)
\notag\\
   &\qquad\qquad\qquad{}
   +(1+y\cdot\partial+2\Delta)L_\mu^a(y)\cdot
   \frac{\delta}{\delta L_\mu^a(y)}\Psi_{-1}(x)
   \Biggr],
\label{eq:(A10)}
\end{align}
which is the first of~Eq.~\eqref{eq:(4.7)}. The rest of~Eq.~\eqref{eq:(4.7)}
can be derived in the same way.

\section{Derivation of~Eq.~\eqref{eq:(2.12)}}
\label{sec:B}
The purpose of this appendix is to derive Eq.~\eqref{eq:(2.12)} which relates
the correlation functions of~$S_\tau$ to those of~$S_{\tau_0}$. As a preparation,
we first show
\begin{align}
   &\int[d\psi d\Bar{\psi}]\,
   \hat{s}^{-1}
   \left(\exp\left\{\int d^Dx\,
   \left[\Bar{J}(x)\psi(x)+\Bar{\psi}(x)J(x)\right]\right\}\right)
\notag\\
   &\qquad{}
   \times
   \exp\left\{i\int d^Dx\,
   \left[\Bar{\psi}(x)-\Bar{\Psi}(x)\right]
   \left[\psi(x)-\Psi(x)\right]\right\}
\notag\\
   &=\exp\left\{
   \int d^Dx\,
   \left[\Bar{J}(x)\Psi(x)+\Bar{\Psi}(x)J(x)\right]\right\},
\label{eq:(B1)}
\end{align}
where $J$, $\Bar{J}$, $\Psi$, $\Bar{\Psi}$ are arbitrary anticommuting spinors.
Using
\begin{align}
   &\hat{s}^{-1}
   \left\{\exp\left[
   \int\left(\Bar{J}\psi+\Bar{\psi}J\right)
   \right]\right\}
\notag\\
   &=\exp\left(\int\Bar{J}\psi\right)
   \exp\left(-i\int\frac{\overleftarrow{\delta}}{\delta\psi}
   \frac{\overrightarrow{\delta}}{\delta\Bar{\psi}}
   \right)
   \exp\left(\int\Bar{\psi}J\right)
\notag\\
   &=\exp\left[
   -i\int\Bar{J}J
   +\int\left(\Bar{J}\psi+\Bar{\psi}J\right)
   \right],
\label{eq:(B2)}
\end{align}
we obtain the Gaussian functional integral as
\begin{align}
   &\int[d\psi d\Bar{\psi}]\,
   \hat{s}^{-1}
   \left\{\exp\left[
   \int\left(\Bar{J}\psi+\Bar{\psi}J\right)\right]\right\}
   \cdot
   \exp\left[
   i\int\left(\Bar{\psi}-\Bar{\Psi}\right)
   \left(\psi-\Psi\right)
   \right]
\notag\\
   &=\int[d\psi d\Bar{\psi}]\,
   \exp\left\{
   \int\left[-i\Bar{J}J+\Bar{J}\psi+\Bar{\psi}J
   +i\left(\Bar{\psi}-\Bar{\Psi}\right)\left(\psi-\Psi\right)
   \right]
   \right\}
\notag\\
   &=\int[d\psi d\Bar{\psi}]\,
   \exp\left[
   i\int\left(\Bar{\psi}-\Bar{\Psi}-i\Bar{J}\right)
   \left(\psi-\Psi-iJ\right)
   +\int\left(\Bar{\Psi}J+\Bar{J}\Psi\right)\right]
\notag\\
   &=\exp\left[
   \int\left(\Bar{\Psi}J+\Bar{J}\Psi\right)
   \right].
\label{eq:(B3)}
\end{align}
This proves Eq.~\eqref{eq:(B1)}. Differentiating Eq.~\eqref{eq:(B1)} with
respect to~$J$ and~$\Bar{J}$ multiple times, and setting~$J=\Bar{J}=0$, we
obtain
\begin{align}
   &\int[d\psi d\Bar{\psi}]\,
   \hat{s}^{-1}
   \left[
   \psi(x_1)\dotsb\psi(x_n)\Bar{\psi}(y_1)\dotsb\Bar{\psi}(y_n)
   \right]
\notag\\
   &\qquad{}
   \times
   \exp\left\{
   i\int d^Dx\,
   \left[\Bar{\psi}(x)-\Bar{\Psi}(x)\right]
   \left[\psi(x)-\Psi(x)\right]\right\}
\notag\\
   &=\Psi(x_1)\dotsb\Psi(x_n)\Bar{\Psi}(y_1)\dotsb\Bar{\Psi}(y_n).
\label{eq:(B4)}
\end{align}

Before proceeding to derive Eq.~\eqref{eq:(2.12)}, let us make a little detour
giving an alternative expression for~Eq.~\eqref{eq:(B1)}. By functional
integration by parts, we can move $\hat{s}^{-1}$ to the right and rewrite the
left-hand side of~Eq.~\eqref{eq:(B1)} to obtain
\begin{equation}
   \int[d\psi d\Bar{\psi}]\,
   \exp\left[
   \int\left(\Bar{J}\psi+\Bar{\psi}J\right)\right]
   \cdot\hat{s}^{-1}
   \exp\left[
   i\int\left(\Bar{\psi}-\Bar{\Psi}\right)
   \left(\psi-\Psi\right)\right]
   =\exp\left[\int\left(\Bar{J}\Psi+\Bar{\Psi}J\right)
   \right].
\label{eq:(B5)}
\end{equation}
This implies
\begin{equation}
   \hat{s}^{-1}
   \exp\left[
   i\int\left(\Bar{\psi}-\Bar{\Psi}\right)\left(\psi-\Psi\right)
   \right]
   =\prod_x
   \left[\delta\left(\psi(x)-\Psi(x)\right)
   \cdot
   \delta\left(\Bar{\psi}(x)-\Bar{\Psi}(x)\right)
   \right].
\label{eq:(B6)}
\end{equation}
In Sect.~\ref{sec:3}, we use this formula to derive Eq.~\eqref{eq:(3.10)}.

Let us finish deriving Eq.~\eqref{eq:(2.12)}. Using Eqs.~\eqref{eq:(2.1)}
and~\eqref{eq:(B4)}, we obtain
\begin{align}
   &\left\langle
   \hat{s}^{-1}\psi(x_1)\dotsb\Bar{\psi}(y_1)\dotsb\right\rangle_{S_\tau}
\notag\\
   &=\int[d\psi d\Bar{\psi}]\,
   \hat{s}^{-1}
   \left[\psi(x_1)\dotsb\Bar{\psi}(y_1)\dotsb\right]
   \int[d\psi'd\Bar{\psi}']\,(\hat{s}')^{-1}e^{S_{\tau_0}[\psi',\Bar{\psi}',L']}
\notag\\
   &\qquad{}
   \times
   \exp\bigg\{
   i\int d^Dx\,
   \left[
   \Bar{\psi}(x)-e^{[(D-1)/2](\tau-\tau_0)}\Bar{\psi}'(t-t_0,e^{\tau-\tau_0}x)
   \right]
\notag\\
   &\qquad\qquad\qquad\qquad\qquad{}
   \times
   \left[\psi(x)-e^{[(D-1)/2](\tau-\tau_0)}\psi'(t-t_0,e^{\tau-\tau_0}x)\right]
   \bigg\}
\notag\\
   &\stackrel{\eqref{eq:(B4)}}{=}
   \int[d\psi'd\Bar{\psi}']\,
   (\hat{s}')^{-1}
   e^{S_{\tau_0}[\psi',\Bar{\psi}',L']}
\notag\\
   &\qquad\qquad{}
   \times
   e^{[(D-1)/2](\tau-\tau_0)}\psi'(t-t_0,e^{\tau-\tau_0}x_1)
   \dotsb e^{[(D-1)/2](\tau-\tau_0)}\Bar{\psi}'(t-t_0,e^{\tau-\tau_0}y_1)
   \dotsb
\notag\\
   &=e^{2n[(D-1)/2](\tau-\tau_0)}
   \left\langle\hat{s}^{-1}
   \psi(t-t_0,e^{\tau-\tau_0}x_1)
   \dotsb
   \Bar{\psi}(t-t_0,e^{\tau-\tau_0}y_1)\dotsb\right\rangle_{S_{\tau_0}}.
\label{eq:(B7)}
\end{align}
This is~Eq.~\eqref{eq:(2.12)}.



%\bibliographystyle{ptephy}
%\bibliography{sample}
%
% once the .bbl file has been generated then place the text in your article.

%% \vspace{0.2cm}
%% \noindent
%% For references, note how to include DOI information from examples below. 

%This is added by T. Yoneya (editor-in-chief) on 2020/07/09.

\let\doi\relax

%without this code before the command "\begin{thebibliography}{}" , an error will be %flagged. When the bibliography is provided as separate .bib file, then this code %should be placed above the commands "\bibliographystyle{}" and "\bibliography{}" %inside the main TeX file. 

%% \begin{thebibliography}{9}

%% \bibitem{1}
%% J. P.~Blaizot, and E.~Iancu, Phys. Rep. {\bf 359}, 355 (2002).
%% \doi{https://doi.org/10.1016/S0370-1573(01)00061-8}

%% \bibitem{2}
%% M.~Gyulassy, and L.~McLerran, Nucl.\ Phys.\  A {\bf 750}, 30 (2005). \\ \doi{https://doi.org/10.1016/j.nuclphysa.2004.10.034}

%% \bibitem{3}
%% S.~Aoki et al. [JLQCD Collaboration], Phys. Rev. D 72, 054510 (2005). \\
%% \doi{https://doi.org/10.1103/PhysRevD.72.05451}

%% \bibitem{4}
%% S.~Alekhin, A.~Djouadi, and S.~Moch, Phys. Lett. B 716, 214 (2012) [arXiv:1207.0980 [hep-ph]]. \doi{https://doi.org/10.1016/j.physletb.2012.08.024}

%% \end{thebibliography}

%\bibliographystyle{ptephy}
%\bibliography{anomaly}

\end{document}